# Graphene/Polyelectrolyte Layer-by-Layer Coatings for Electromagnetic Interference Shielding

*Cristina Vallés[†]\*, Xiao Zhang[‡], Jianyun Cao[†], Fei Lin[†], Robert J. Young[†], Antonio Lombardo[‡], Andrea C. Ferrari[‡], Laura Burk[§], Rolf Mülhaupt[§] and Ian A. Kinloch[†]\**

[†] School of Materials and National Graphene Institute, University of Manchester, Oxford Road, Manchester, M13 9PL, UK

[‡] Cambridge Graphene Centre, University of Cambridge, 9 JJ Thomson Avenue, Cambridge CB3 0FA, UK

[§] Institute for Macromolecular Chemistry and Freiburg Materials Research Center FMF, Stefan-Meier-Straße, Freiburg, Germany

ABSTRACT. Electromagnetic interference (EMI) shielding coating materials with thicknesses in the microscale are required in many sectors, including communications, medical, aerospace and electronics, to isolate the electromagnetic radiation emitted from electronic equipment. We report a spray, layer-by-layer (LbL) coating approach to fabricate micron thick, highly-ordered and electrically-conductive coatings with exceptional EMI shielding effectiveness (EMI SE

≥4830 dB/mm), through the alternating self-assembly of negatively-charged reduced graphene oxide (RGO) and a positively-charged polyelectrolyte (PEI). The microstructure and resulting electrical properties of the $(PEI/RGO)_n$ LbL structures are studied as function of increasing mass of graphene deposited per cycle (keeping the PEI content constant), number of deposited layers ($n$), flake diameter and type of RGO. A strong effect of the lateral flake dimensions on the electrical properties is observed, which also influences the EMI SE. A maximum EMI SE of 29 dB is obtained for a 6 μm thick $(PEI/RGO)_{10}$ coating with 19 vol.% loading of reduced electrochemically-exfoliated graphene oxide flakes with diameters ~3μm. This SE performance exceeds those previously reported for thicker graphene papers and bulk graphene/polymer composite films with higher RGO or graphene nanoplatelets contents, which represents an important step towards the fabrication of thin and light-weight high-performance EMI shielding structures.

INTRODUCTION

New environmental pollution problems[1] are emerging due to the rapid development of electronic technologies, including telephones, computers and radios. Electronics and their components with higher power, smaller size and faster operative speed emit unwanted electromagnetic waves, causing electromagnetic interference (EMI) between different electronic devices with detrimental impact on their performance.[2-5] Recently, electrically conductive polymer composites have raised much attention for EMI shielding applications[6-17] because of their light weight, resistance to corrosion, flexibility, good processability and low cost compared to conventional metal-based materials.[2] High electrical conductivity and connectivity of the conductive fillers in polymer composites are key factors to optimize their EMI shielding performance.[6, 7, 9, 12, 13] Carbon-based materials have been investigated as conductive fillers to fabricate composite materials for EMI shielding because they offer a combination of high electrical conductivity, excellent mechanical properties, light weight, flexibility and large aspect ratios.[6-17] In particular, graphene emerges as highly promising for EMI shielding applications due to its excellent in-plane electrical conductivity ($\sim 4.5 \times 10^4$ S/cm)[14, 18-20] and easy processability.[21] Graphene-based polymer composites, foam structures, aerogels, thin films and papers have already been investigated as light-weight EMI shielding materials.[2, 22-30] An EMI SE $\sim$500 dB·cm$^3$/g was reported for a graphene foam composite with a density $\sim$0.06 g/cm$^3$,[2] and $\sim$33780 dB·cm$^2$/g for ultralight cellulose fiber/thermally reduced GO hybrid aerogels, with a density as low as 2.83 mg/cm$^3$.[22] More recently, EMI SE of 21.8 dB have been reported for thermoplastic polyurethane/reduced graphene oxide composite foams with only 3.17 vol.% of RGO[27], whereas EMI SE of 48.56 dB has been found for 3D-interconnected graphene aerogels decorated with cobalt ferrite nanoparticles and ZnO nanorods[28]. Graphene papers with thicknesses between 12.5 and 470 μm[23, 24] and ~30-60 μm thick graphene/polymer composites[25, 26] were also investigated as EMI shielding materials. Wan *et al*. reported EMI SE of up to ~52.2 dB for iodine doped and ~47 dB for 12.5 μm thick undoped reduced GO papers[23], whereas EMI SE ~55.2 dB was reported for 0.47 mm thick multilayer films composed of highly ordered nitrogen-doped graphene.[24] More recently, EMI SE up to 26 dB have been reported for transparent films based on graphene nanosheets and silver nanowires[29], whereas values of 58.5 dB have been found for 6.6 μm-thick nitrogen-doped graphene films[30]. EMI SE ~14 dB was found for 0.03-0.05 mm thick free-standing graphene/thermoplastic polyurethane (TPU) composite films with 0.12 vol. fraction of liquid-phase-exfoliated pristine graphene nanosheets,[25] whereas ~27 dB was achieved for 0.35 mm thick multilayer graphene/polymer composite with 0.60 vol. fraction of graphene.[26] Despite being required in sectors such as medical, aerospace or electronics, micron-thick graphene-based coatings with reduced graphene contents have not been exploited yet for EMI, due to the difficulty to overcome the ~20 dB target required for commercial applications[2] with such low amounts of graphene (relative to most of the previously described works).

Layer-by-layer (LbL) assembly of oppositely-charged polyelectrolytes and inorganic particles can be used to fabricate highly-ordered multilayer nanostructured films and coatings with thickness in the nano- or micro- scale.[31] This coating method offers important advantages, such as uniform coverage of the substrate and controllable film/coating thicknesses.[31] Another advantage is its great versatility, as it can be used with a wide variety of fabrication processes, including dip-, spray- or spin-coating.[31] A few reports on the successful preparation of GO-polymer self-assembled multilayer structures using LbL approaches with remarkable gas barrier properties can already be found in the literature[32-34]. Transparent coatings/films prepared by LbL electrophoretic deposition of RGO,[35] or by chemical vapor deposition (CVD) graphene alternated with spin coated PMMA[36] were fabricated as transparent EMI shielding materials. In these works,[35, 36] the LbL coatings were manufactured using dip-coating or electrophoretic deposition due to their ability to produce nanometer thick coatings with low graphene contents and hence high transparency.[35-37] However, for the majority of EMI applications, transparency is not required, allowing more scalable routes, such as spray coating, to be employed with high concentrations of graphene ($\geq$ 10 wt.%). Thus a detailed evaluation of how the microstructure and electronic properties develop in LbL multi-layered structures with increasing amounts of graphene is needed to optimize the properties of the coatings, while minimizing thickness, weight and amount of graphene employed.

Here, we develop a spray LbL coating technique to fabricate thin (up to 30 μm thick), highly-organized and electrically-conductive graphene based coatings on a substrate, alternating negatively-charged reduced graphene oxide flakes and positively charged polyelectrolyte (PEI) to render (RGO/PEI)$_n$ coatings ($n$ = number of deposition cycles), as schematically represented in Figure 1. We show how the microstructure, level of orientation of the graphene flakes and electrical properties of the LbL coatings develop as a function of the amount of reduced graphene oxide deposited per cycle, while keeping the amount of PEI and $n$ constant, and we relate it with their EMI shielding effectiveness. An outstanding maximum EMI SE of 29 dB was obtained for a 6 μm thick (PEI/RGO)$_{10}$ coating with 19 vol.% loading of reduced electrochemically-exfoliated graphene oxide flakes with diameters ~3μm, making our coatings potentially promising for EMI shielding applications.

RESULTS AND DISCUSSION

**Microstructure of the (PEI/TRGO)$_n$ LbL coatings**

Multi-layered graphene coatings were prepared using a LbL approach by alternating self-assembly of negatively charged thermally reduced GO (TRGO) and a positively-charged polyelectrolyte (PEI), $n$ times on a PET substrate, as represented schematically in Figure 1. (Zeta potential revealed negative charges of about -50 mV and -40 mV for 0.5 mg/mL dispersions of 100μm-TRGO and 20μm-TRGO, respectively, in ethanol, whereas a positive charge of ~12 mV was found for a 10 mg/mL PEI/H$_2$O solution). The microstructure of the top surface and transverse section of the LbL (PEI/TRGO)$_n$ coatings is studied by SEM (details in the Experimental Section) and the micrographs are shown in Figure 2. The top surface has wrinkles (Figures 2a,b) similar to the topology of the TRGO flakes (TEM micrographs Figure S1, SI). This is consistent with a LbL formation where the particles dictate the morphology of the deposited polymer layers. SEM micrographs of the transverse sections of the LbL coatings in Figures 2c,d reveal that the spray-on approach produces highly-ordered and well-defined multi-layered microstructures.

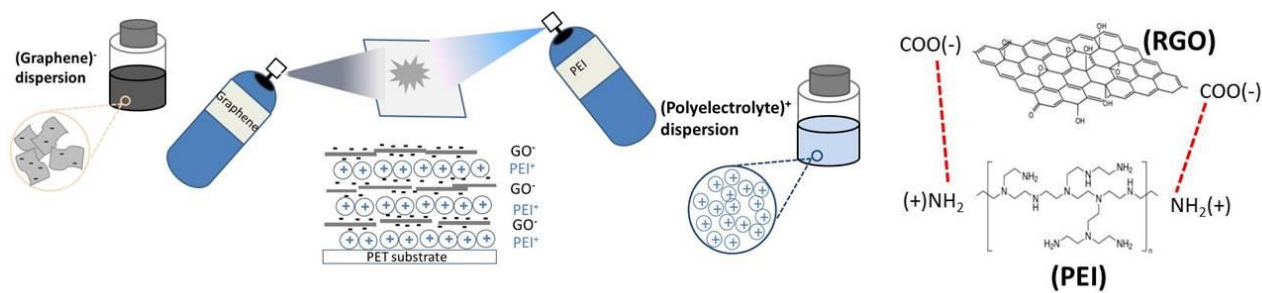

**Figure 1.** Schematic of the spray coating approach used to fabricate the LbL coatings through alternating the absorption of negatively-charged TRGO and a positively-charged polyelectrolyte (PEI) on a PET substrate, rendering (PEI/TRGO)$_n$ coatings ($n$ = number of bi-layers).

To further characterize the layered microstructure of these coatings, polarized Raman spectroscopy was used. Polarized Raman spectroscopy has been previously employed to evaluate and quantify the orientation of graphene oxide flakes in a PVA composite.[38] Following this procedure described by Li *et al.*[38], the orientation of the TRGO flakes in the LbL coatings prepared here was evaluated using a backscattering geometry and obtaining spectra with the polarized laser beam aligned either perpendicular (in the z-direction) or parallel (in the x-direction) to the surface of the specimen as shown in Figure 3. The specimen was rotated relative to its axes and an analyzer was used with the scattered light. A VV (vertical/vertical) combination of incident and scattered polarization was employed in which the directions of the incident and scattered polarization were the same, and the change in the intensities of the D band, $I_D$, with specimen orientation was monitored. The variation of $I_D$ for the studied coatings for the two laser beam directions with the relative orientations of the incident laser polarization are shown in Figure 3. (The intensities have been normalized in each case for a maximum intensity of unity). In the case where the laser beam is parallel to the z-axis (*i.e.*, perpendicular to the plane of the coating, $\Phi_z$) in Figure 3 it can be seen that the $I_D$ remains approximately constant as the specimen is rotated. In contrast, when the laser beam is parallel to the x-axis ($\Phi_x$, Figure 3) the $I_D$ is a maximum when the direction of laser polarization is parallel to the specimen edge plane and a minimum when it is perpendicular to the edge.[39] This is consistent with the in-plane alignment of the TRGO layers in the LbL coating structure seen in the SEM micrograph in Figure 2.

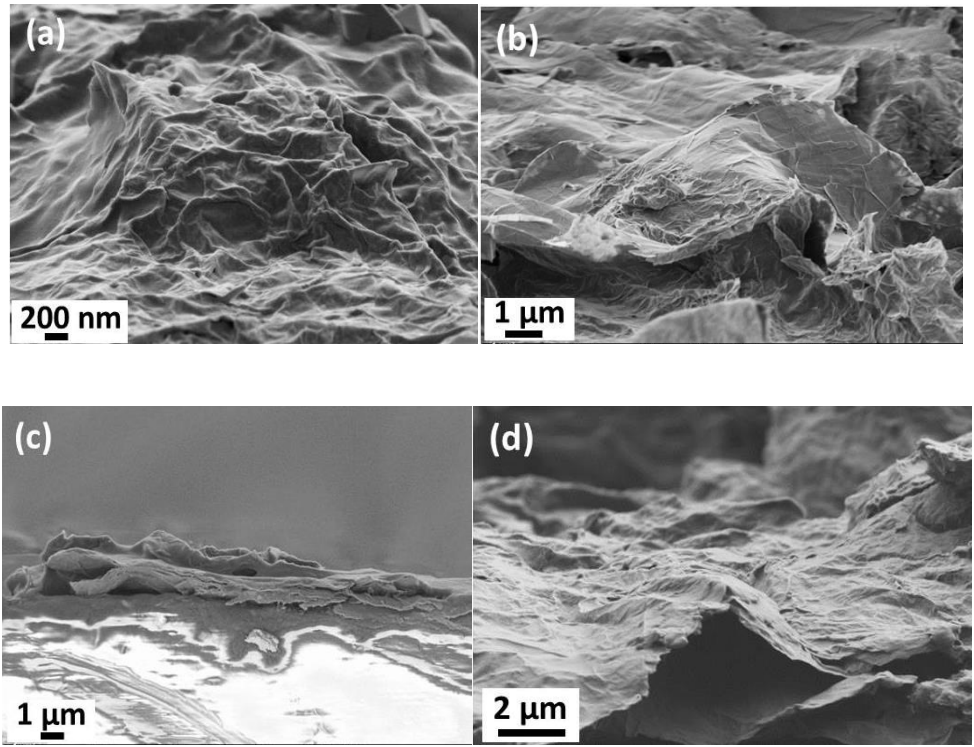

**Figure 2.** SEM images of the surface morphology of (PEI/20μm-TRGO)$_5$ (a) and (PEI/100μm-TRGO)$_5$ (b). SEM images of the transversal section of (PEI/20μm-TRGO)$_5$ LbL coating (c, d). (TRGO loading = 14 vol.%).

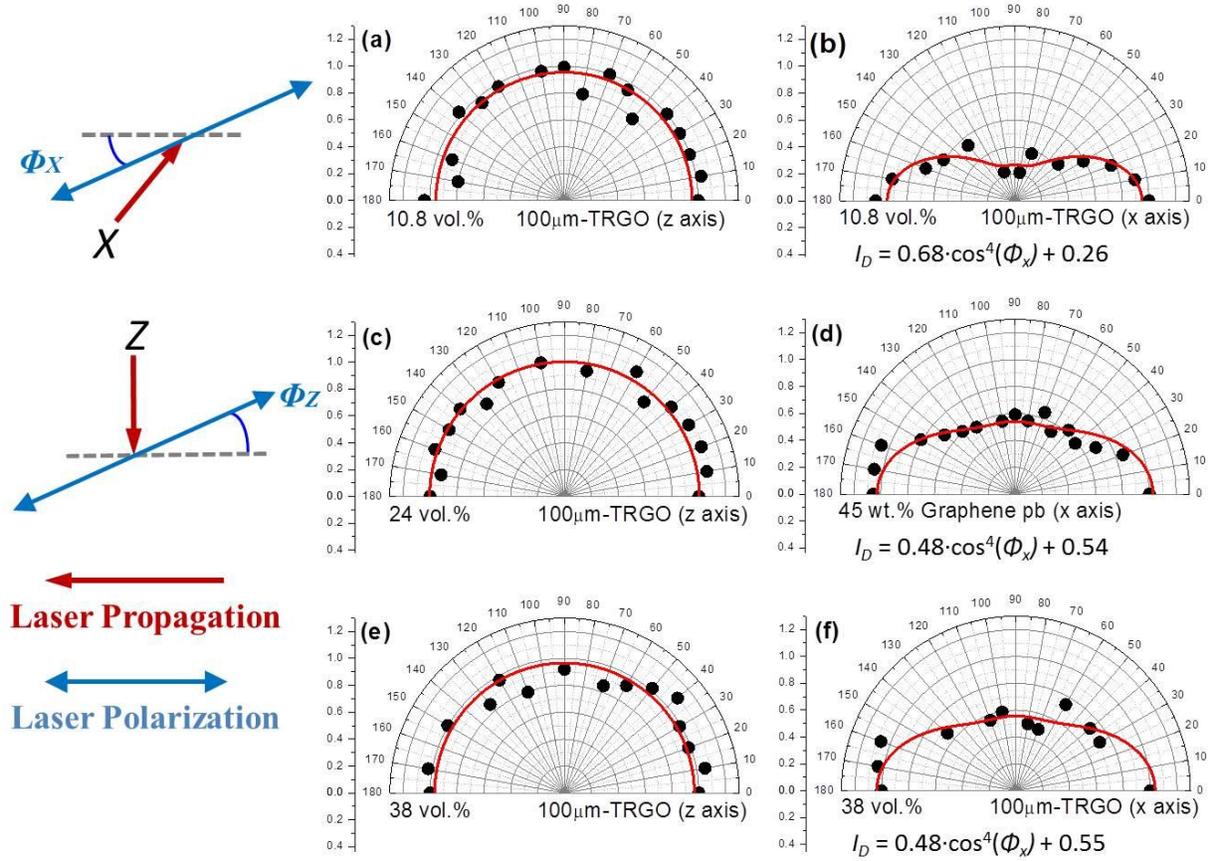

**Figure 3.** Variation of $I_D$ as a function of $\Phi z$ (a, c, e) and $\Phi x$ (b, d, f) for (PEI/100μm-TRGO)$_{10}$ coatings with different graphene loadings. Schematic description of the *x*, *y* and *z* directions in the coating specimens.

A fitting of the data plot in Figure 3 follows a function on the form:

$$I_D = A \cdot \cos^4 \Phi_x + B \qquad \text{Equation (1)}$$

with A = 0 and B = 1 for completely randomly oriented flakes, and A = 1 and B = 0 for a perfect alignment of the flakes. A higher level of orientation (*i.e.* a higher A and a lower B, see Figure 3) is detected in coatings with 10.8 vol.% graphene than with 24 and 38 vol.%, for the same *n*. This may be due to a combination of more overlap between the flakes and the formation of small agglomerates of graphene at higher loadings. The former is related to the onset of complete in-plane coverage, meaning that the 24 and 38 vol.% coatings have similar levels of orientation (see later), whilst the latter is consistent with Li's work[38] where a decreasing orientation with increasing GO loading was observed in PVA composites and attributed to aggregation.

All samples have a similar crumpled surface morphology (Figures 2a,b) and multi-layered structure (transverse surface, Figures 2c,d). Their thicknesses (shown in the SI, Figure S4) depended strongly on loading and $n$, as well as on the flakes lateral dimensions. (The graphene loading, *i.e.* the amount of graphene deposited per cycle can be easily controlled by modifying the deposited volume of a dispersion of graphene with a known concentration). A linear increase of the overall thickness with increasing loading is observed for constant $n$. At very low loadings of graphene (< 0.027 vol.%) similar thicknesses were observed for 20 μm and 100 μm diameter flakes (named as 20μm-TRGO and 100μm-TRGO, respectively) based coatings as expected from depositing TRGO flakes with similar thicknesses, whereas the coatings fabricated with the 100μm-TRGO are thicker at higher graphene loadings, probably due to increased overlap of the large flakes relative to the small flakes. The thicknesses of the coatings increase with $n$ for constant graphene loading.

**Electrical properties**

In order to understand how the electrical conductance develops in multi-layer microstructures, the electrical conductivity of the LbL coating structures fabricated was evaluated with increasing graphene loading per cycle and $n$. Log-log plots of conductivity as a function of frequency for (PEI/20μm-TRGO)$_n$ and (PEI/100μm-TRGO)$_n$ at increasing loading of graphene for $n = 2$ and $n = 10$ are shown in Figure 4. Figure 4 shows the behaviour typical of 2 phase systems, *i.e.* a resistive component (graphene) and a capacitive component (PEI, *i.e.* a dielectric material). Below the percolation threshold, the conductivity depends linearly on frequency, as characteristic of dielectric behaviour, and typical of insulating materials.[40] Above percolation the conductivity plateaus up to a critical frequency at which the behaviour reverted to a dielectric behaviour. This indicates the formation of conductive paths within the matrix and a dominance of the resistive component, combined with a dielectric behaviour at higher frequencies where the capacitive component (*i.e.* the polyelectrolyte and network junctions) dominates. At high graphene loadings (> 13.5 vol.% for *n* = 2 and > 24.3 vol.% for *n* = 10) the conductivity is independent of the frequency across the entire frequency range. This is typical of an electrically-conductive material,[40] similar to that observed for a pure graphene paper (*i.e.* a free-standing paper prepared by vacuum filtration of a dispersion of the as-prepared graphene), also shown in Figure 4.

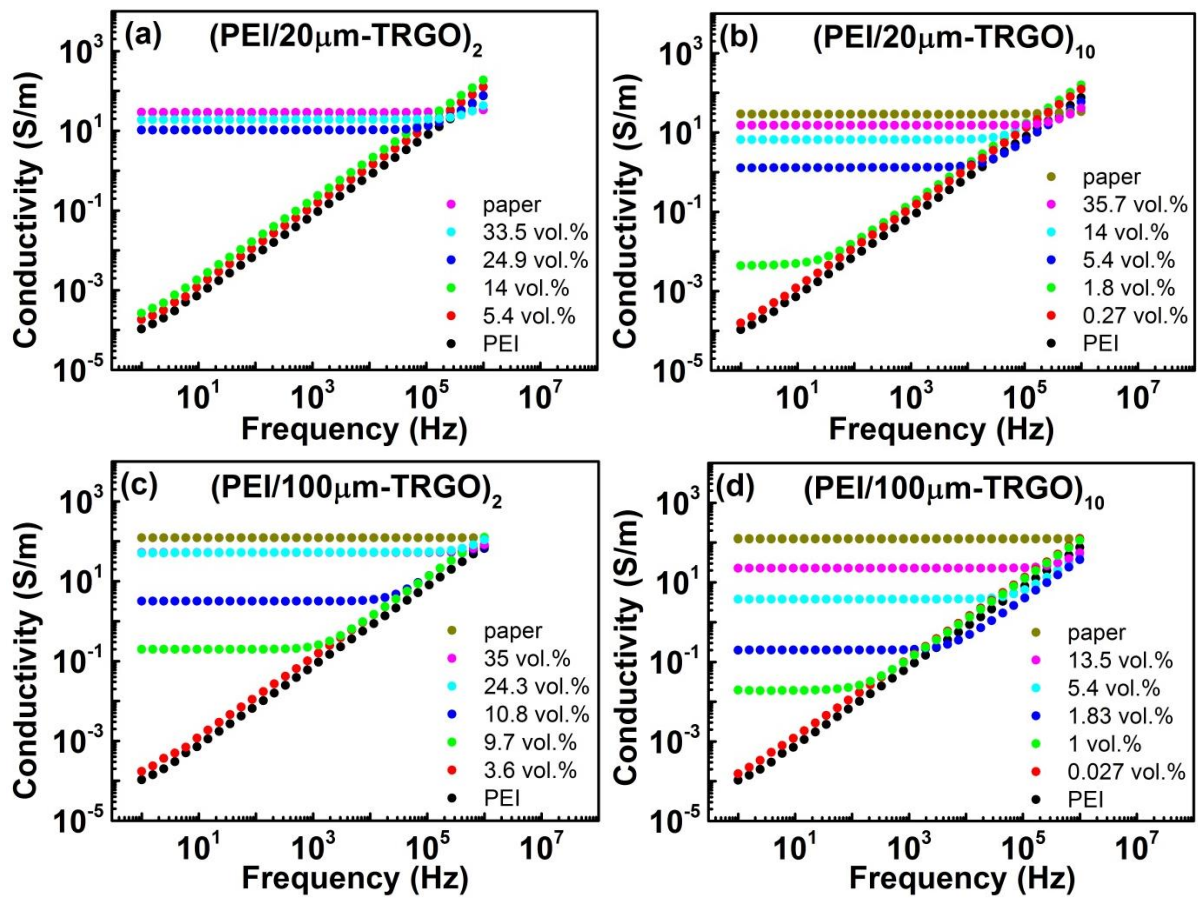

**Figure 4.** Log-log plots of conductivity as a function of frequency for (PEI/20µm-TRGO)$_n$ and (PEI/100µm-TRGO)$_n$ LbL coatings ($n = 2$, $n = 10$) for different graphene loadings.

*Percolation threshold*

The percolation threshold depends on the flakes lateral dimensions and $n$. For the (PEI/20µm-TRGO)$_n$ systems the percolation is between 14.05 and 24.86 vol.% of graphene for $n = 2$, and between 0.3 and 1.8 vol.% for $n = 10$. For LbL coatings with 100µm-TRGO this is in the ranges of 3.6-9.8 vol.% and 0.2-1.19 vol.% for $n = 2$ and $n = 10$, respectively. Figure 5 shows the conductivities of the LbL structures at 1Hz as a function of graphene loading for the two graphene types studied here at $n = 2,10$. These show *S*-shape curves typical of percolated systems,[40] suggesting that classical percolation theory[41] could be used to calculate the percolation threshold ($P_c$), defined as the critical/minimum amount of filler required to transform a non-conductive material into a conductive one. Assuming that the fillers are distributed randomly within the polymer matrix, the conductivity can be described as:[40]

$$\sigma = \sigma_0 (P - P_c)^t \qquad \text{Equation (2)}$$

where $P$ is the fraction of filler in the LbL structure, $P_c$ the percolation threshold, $t$ the conductivity exponent, $\sigma_0$ is the electrical conductivity of the matrix and $\sigma$ the electrical conductivity of the LbL system.

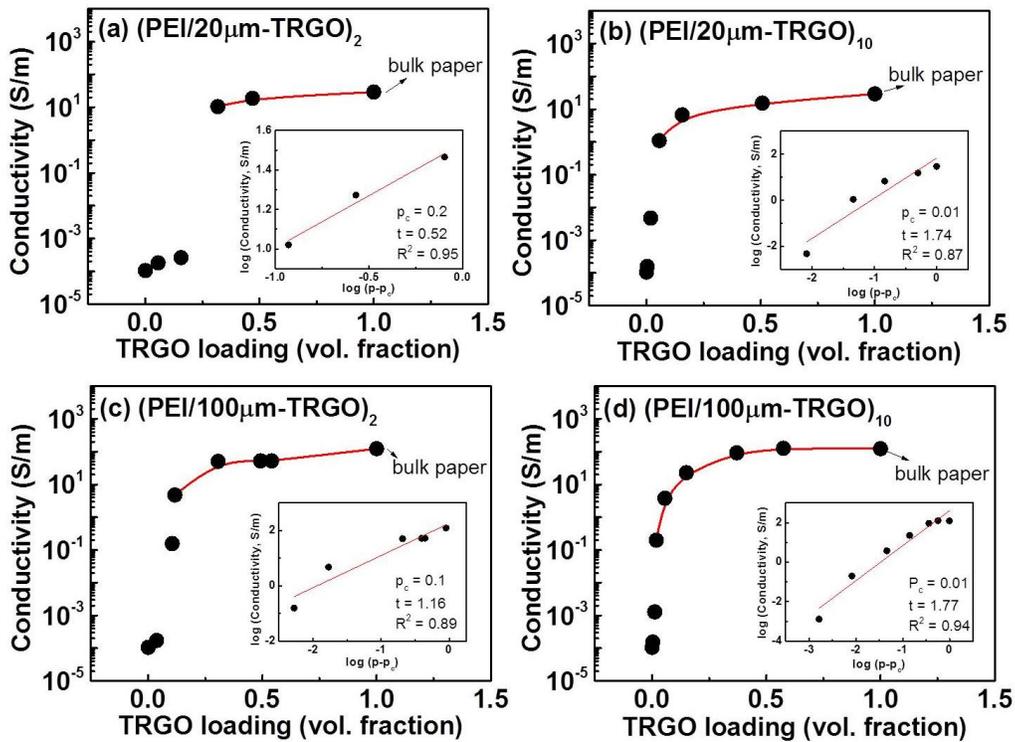

**Figure 5.** Semi-log plot of conductivity as a function of graphene loading for the LbL coated systems: (a) (PEI/20μm-TRGO)$_2$; (b) (PEI/20μm-TRGO)$_{10}$; (c) (PEI/100 μm-TRGO)$_2$; (d) (PEI/100μm-TRGO)$_{10}$. (The inserts are log-log plots of conductivity as a function of $P$-$P_c$, determined from the vol. fraction data).

From Figure 5, $P_c = 0.20$ for $n = 2$ and $P_c = 0.01$ were determined for $n = 10$ for (PEI/20μm-TRGO)$_n$. Linear fittings of log (*Conductivity*) vs. log ($P$-$P_c$) data (insert in Figure 5) give $t = 0.52$ and 1.74 for $n = 2$, 10, respectively. Similarly, for the (PEI/100μm-TRGO)$_n$, $P_c = 0.10$ for $n = 2$ and $P_c = 0.01$ for $n = 10$ were found, resulting in $t = 1.16$ and 1.77, respectively.

Lower percolation thresholds were found for 100μm- than for 20μm-TRGO for identical graphene loading and *n*. Also, the percolation threshold for $n = 10$ is lower than for $n = 2$ (for identical loading of graphene and independent on the lateral dimensions). The percolation thresholds are summarized in Table 1.

**Table 1.** Experimental (determined from the data shown in Fig. 5) and theoretically-calculated percolation thresholds and coverage graphene loadings for the LbL coatings.

| Graphene | *n* | Exp. Perc. Threshold | [1]Theor. Perc. Threshold | [2]Exp. Coverage | [3]Theor. Coverage |
|---|---|---|---|---|---|
| 20μm-TRGO | 2 | 0.20 | 0.0013-0.0037 (vol. fraction) | 0.24 (vol. fraction) | 0.002-0.0059 (vol. fraction) |
| | 10 | 0.01 | | | |
| 100μm-TRGO | 2 | 0.10 | | | |
| | 10 | 0.01 | | | |

[1]according to classical percolation models[42] (fraction of graphene which correspond to an area occupation of 68%); [2]corresponding to the transition from *n*-dependency to *n*-independency; [3]calculations shown in SI (fraction of graphene corresponding to an area occupation of 1).

Pike and Seager [42] studied a two dimensional random-lattice site system at a critical area fraction of ~0.68 using Monte Carlo calculations and assuming particles to be uniform discs with hard cores. If we consider the TRGO flakes as uniform discs deposited on a substrate (in one deposition cycle) with overlapping, the percolation should happen theoretically at 0.0013–0.0037,[42] which corresponds to the vol. fraction of graphene calculated for one deposition cycle with the graphene occupying 68 % of the surface of the substrate (see SI). The theoretical values are lower than those derived from our experiments, which we relate to the presence of defects on the TRGO flakes coming from the reduction process, the presence of gaps possibly associated to a non-highly homogeneous distribution of the flakes, as well as to the fact that TRGO flakes are not discs with uniform diameters. The wrinkled nature of TRGO flakes (TEM, SI) may also contribute to this discrepancy.

For a percolated system, $t$ depends on the dimensionality of the composites. Theoretically, $t$ should vary from ~1.6 to 2.0 for 3D-percolated systems and from 1 to ~1.6 for 2D.[40, 43] Applying the classical percolation theory[42] to our data, we can say that our LbL coatings show a 2D-percolation mechanism, with $t$ increasing with $n$, which suggests an evolution from 2D to 3D-percolated systems from $n = 2$ to $n = 10$. In addition, higher values of $t$ are found for larger flakes, as expected from percolation theory.[40]

Zhao *et al.*[44] studied percolation in graphitic conductive networks embedded in polymer composites and reported $t = 2.40$-$6.92$, corresponding to 3D-percolation. Du *et al.*[45] reported $t = 1.08$ for a 2D-percolated segregated graphene network in bulk high density polyethylene composites processed by hot-press, which suggests that the level of orientation/organization of the fillers in the matrix must be defining the dimensionality of conduction. We thus assign the level of organization achieved in our LbL coatings (evidenced by SEM and polarized Raman) as responsible for the 2D conduction mechanism found here. The differences between theory and experimental data, and the discrepancies between different reported experimental systems, must be related to additional factors influencing the percolation, such as additional dimensional aspects of the filler network, presence of agglomerates, polymer-filler interaction, level of organization/orientation, *etc*.

*Development of the conductance in LbL structures above percolation*

Above percolation, the conductivities increase with graphene loading and *n* up to a maximum limit, Figure 6. The *n*-dependence before saturation suggests that when the electrons find defects or gaps in one graphene layer, they tend to hop to the next level (from $n = 1$ to $n = 2$, …, as schematically represented in Figure 6), leading to enhancement of overall conductivities (the variation of the conductivities of the coatings with increasing *n* is shown in Figure S6 in the SI). This observation agrees with the small increase in *t* with increasing *n, i.e.* the increasing 3D nature of the ne*twork*. Even with the electrons hopping to contiguous layers to find more conductive paths, the systems behave as 2D percolated systems.

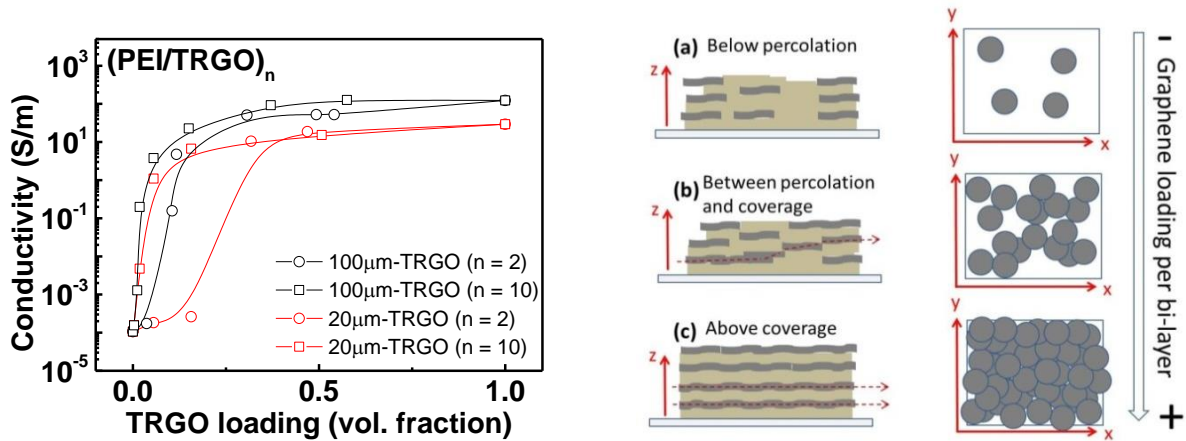

**Figure 6.** Comparison of the conductivity of (PEI/20μm-TRGO)$_n$ and (PEI/100μm-TRGO)$_n$ LbL coatings for different *n* and schematic representation of the three regions found for the electrical behaviour of the LbL coatings with increasing graphene loading: (a) below percolation; (b) between percolation and coverage (conductivities are *n*-dependent); (c) above coverage (conductivities are *n*-independent).

*Coverage*

There is a second critical graphene loading at which conductivities reach a maximum (Figure 6). This maximum cannot be further increased, neither by increasing the graphene loading nor by increasing *n* and corresponds to the conductivity measured for bulk papers composed purely of graphene (*i.e.* the inherent conductivity of the graphene employed for the fabrication of the LbL structures where the inter-flake resistance gives an intrinsic limit).

We attribute the transition from *n*-dependency to *n*-independency to the formation of a continuous network of graphene in one deposition cycle, which we define as *complete coverage* (*i.e.* 100 % occupation of the surface). The experimental coverage loading is found at a vol. fraction ~0.24, significantly higher than the theoretical loading (calculations shown in the SI), similar to what we observed for the percolation and due to the same reasons (the theoretical coverage vol. fraction of graphene is estimated ~0.001-0.0032).

Figure 3 reveals lower orientation at vol. fractions of TRGO $\geq$ 0.24 (*i.e.* above complete coverage), related to the presence of more wrinkles and overlapping between flakes above coverage relative to a situation in which 'isolated' TRGO flakes are deposited with not many connections, overlaps or agglomerates (as represented schematically in Figure 6). All experimental and theoretically-calculated percolation thresholds and graphene loadings are compiled in Table 1. (The % coverage corresponding to the fractions studied here is presented in the SI).

At graphene loadings above complete coverage, interlayer hopping no longer provides paths of higher conductivity. Hence the conductivity becomes independent of *n* and the maximum conductivities of 19 S/m for 34 vol.% of 20μm-TRGO and 127 S/m for 39 vol.% of 100μm-TRGO are found. The maximum conductivity corresponds to the inherent conductivity of graphene paper, highlighting the importance of choosing an electrically conductive graphene, as required for the targeted application, but with sufficient charged groups for LbL formation.

Reaching coverage in one deposition cycle is an important parameter to optimize the electrical properties while minimizing thickness and graphene loading. Within the $n$-dependent region the coatings are typically thicker and less conductive than those fabricated with just one deposition cycle using graphene loadings above coverage. Most of the reported work[46, 47] on multi-layered structures did not achieve coverage in one deposition cycle, they are typically focused on increasing the conductivities by incrementing the number of deposition cycles until saturation. In the works reported by D. G. Wang *et al.*[46] and L. Wang *et al.*[47] electrons were assumed to hop between layers, in agreement with our results below coverage (shown in Figure S6, SI).

The self-assembly of the 100μm-TRGO flakes gives higher conductivities and lower percolation thresholds relative to those found for the self-assembly of the 20μm-TRGO ones for identical $n$, which we relate to the presence of more flake-flake connections and defects in 20μm-TRGO than in 100 μm-TRGO (also observed for the bulk papers). Previously reported simulations and experiments showed that fillers with larger aspect ratios lead to reduced percolation thresholds in systems with higher dimensionalities[48-50].

**EMI shielding**

The coatings with the largest amounts of graphene material and the highest electrical conductivities were then EMI tested. Figure 7 plots the EMI SE for (PEI/TRGO)$_{10}$ LbL structures prepared with increasing graphene in the range of 8-13 Hz, widely used in communication applications such as TV, microwaves and telephones.[51] Figure 7a indicates that the EMI SE is almost constant across the entire range of frequencies. These data reveal that EMI SE increases linearly with increasing graphene loadings for a constant $n$. The data can be linearly fitted, Figure 7b, indicating higher EMI SE with increasing graphene content.

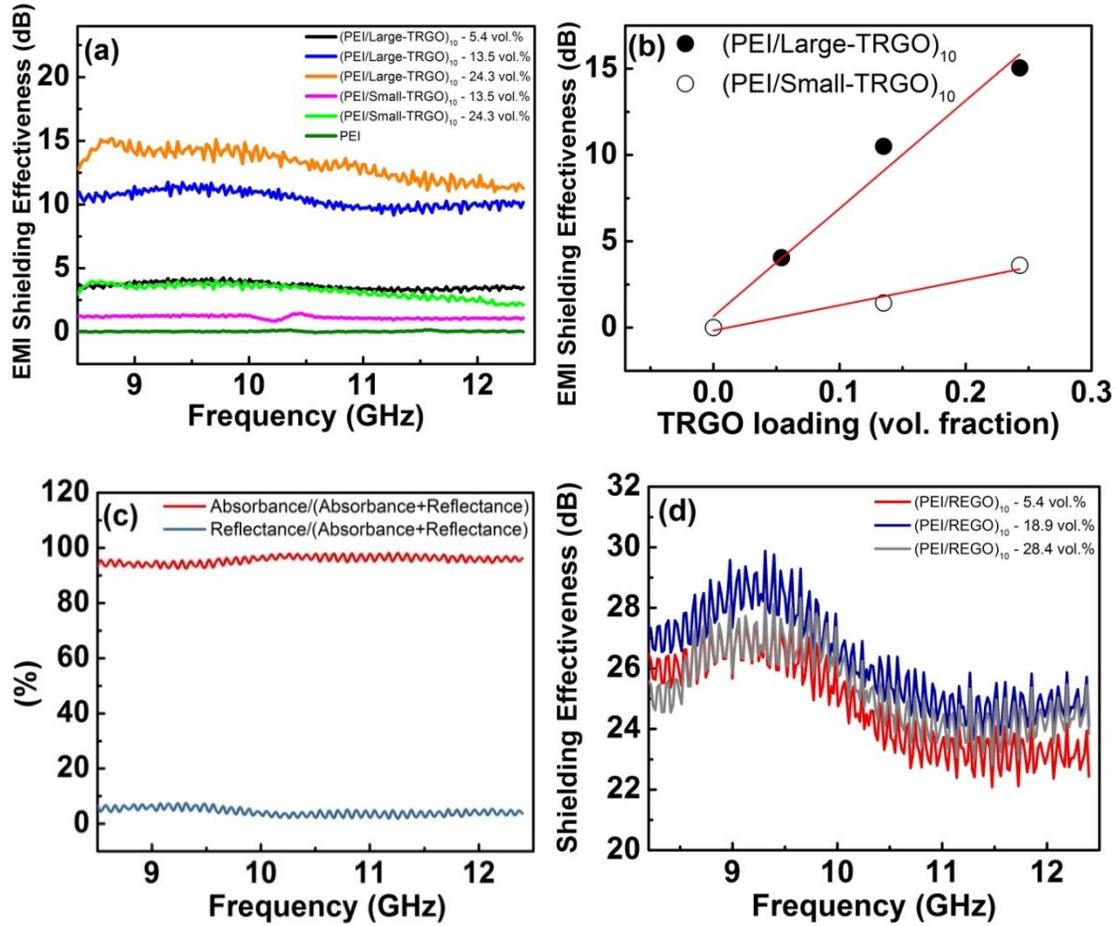

**Figure 7.** (a) Total EMI SE of (PEI/100μm-TRGO)$_n$ and (PEI/20μm-TRGO)$_n$ LbL coating structures; (b) EMI shielding effectiveness vs. vol. fraction of TRGO. The equations obtained from linearly fit these data are: EMI SE (dB) = 14.63·$vol$ ($R^2$ = 0.93), and EMI SE (dB) = 62.47·$vol$ ($R^2$ = 0.97) for (PEI/20μm-TRGO)$_{10}$ and (PEI/100μm-TRGO)$_{10}$, respectively. (c) Absorbance (=Absorbance/(Absorbance+Reflectance)) and Reflectance (=Reflectance/(Absorbance+Reflectance)) contributions. (d) Total EMI SE of (PEI/REGO)$_{10}$ LbL coating structures.

A strong effect of the lateral dimensions of TRGO flakes on the EMI SE is also observed, related to the superior electrical properties of the larger TRGO flakes relative to the smaller ones. These results are in agreement with the work by Wan *et al*.[23] where superior EMI shielding was reported for papers fabricated with larger-size graphene sheets. These authors also attributed this finding to fewer defects and a more conjugated carbon domain size in larger flakes, leading to higher electrical conductivities and, thus, to superior EMI SE.

In order to establish fundamental basis for designing high-performance EMI shielding materials, the mechanism for improved EMI shielding performance was investigated. Figure 7 plots Absorbance (= Absorbance/(Absorbance+Reflectance)) and Reflectance (= Reflectance/(Absorbance+Reflectance)) contributions to the total EMI SE for a (PEI/100μm-TRGO)$_{10}$ coating. This shows a 10% contribution of reflectance and a 90% absorbance, indicating that absorption is the dominant shielding, in agreement with previously reported works[23, 36].

Our LbL coating approach can also be used for other negatively-charged graphene materials, such as hydroiodic acid (HI)-reduced electrochemically exfoliated GO (EGO).[52] The zeta potential technique revealed a negative charge of around -45 mV for a 1 mg/mL aqueous dispersion of EGO and its electrical conductivity is much higher than TRGO. The variation of sheet resistance of LbL coatings with increasing graphene loading is shown in Figure S8 in the SI. The maximum conductivity (*i.e.* minimum sheet resistance) depends on the type of graphene (level of defects and inherent conductivity), which suggests the possibility of improving and tuning the electrical properties of the coatings by using graphene materials with lower amount of defects and, thus, higher conductivities. As a consequence, the total EMI SE of these reduced electrochemically produced graphene based coatings (Figure 7d) is ~5 times higher than TRGO-based coatings with similar graphene content and *n*. In addition, the (PEI/R-EGO)$_n$ coatings are thinner than those prepared using TRGO, which makes these EMI SE better in terms of dB/mm (see Table 2 for comparison with literature data on graphene-based papers and graphene/polymer composite films). We get EMI SE ~29 dB for ~6 μm thick samples (SEM micrograph, Figure S9, SI), These results suggest that our approach can be used for other processable graphene materials (provided they have negatively-charged surfaces), which offers the possibility of tuning electrical conductivity and EMI SE to fit the requirements of emerging applications.

**Table 2.** Comparison between the EMI SE data found in this work and previously reported data for graphene papers and graphene/polymer composite films.

| Material | EMI SE (dB) | Thickness (mm) | Vol. fraction graphene | dB/mm | Reference |
|---|---|---|---|---|---|
| LbL (TRGO/PEI)$_n$ coatings | 15 | 0.03 | 0.30 | 500 | This work |
| LbL (R-EGO/PEI)$_n$ coatings | 29 | ≤0.006 | 0.30 | ≥4830 | This work |
| Graphene paper | 47– 52.2 | 0.0125 | 1 | 3760-4176 | 23 |
| N-doped graphene paper | 55.2 | 0.47 | 1 | 117.4 | 24 |
| Graphene/TPU film | 14 | 0.03 – 0.05 | 0.12 | 280 – 466 | 25 |
| Graphene/epoxy | 27 | 0.04 – 0.06 | 0.6 | 675 – 450 | 26 |
| RGO-EDA | 58.5 | 0.0066 | 1 | 8863 | 30 |

The maximum EMI SE per unit thickness in (PEI/100μm-TRGO)$_{10}$ LbL coating (500 dB/mm) and R-EGO based coating (4830 dB/mm) compare favourably with literature. Wan *et al.*[23] reported EMI SE values of up to ~52.2 dB for iodine-doped graphene papers and ~47.0 dB for undoped graphene papers with thicknesses ~12.5 μm. A similar EMI SE (~55.2dB) was reported for multilayer films of highly-ordered nitrogen-doped graphene with a much higher thickness (0.47 mm),[24] whereas the work by Lin *et al.*[30] reported slightly higher EMI SE values for nitrogen doped graphene papers with higher total amounts of graphene. Thus, our LbL coatings have much better EMI shielding with considerably lower graphene content (Table 2). Our EMI SE is also superior to most previously studied graphene/polymer composites, typically using higher graphene contents.[25, 26] Our results are better than those reported by Song *et al.*[26]

(EMI SE up to ~27 dB) for much thicker (0.35 mm) films of multilayer graphene/epoxy composite containing 0.60 vol. fraction of graphene, as well as better than the EMI SE ~14 dB reported for free-standing conducting graphene/TPU composite films with 0.12 vol. fraction of graphene and thicknesses between 0.03 and 0.05 mm.[25] Thin coatings/films prepared by LbL electrophoretic deposition of RGO[35] or by chemical vapor deposition (CVD) graphene alternated with spin coated PMMA[36] were fabricated as transparent EMI shielding materials. Although they rendered very high values for dB/mm (2-3 orders of magnitude higher than the values found here), the absolute EMI SE values are limited by the transparency required in some sectors. The work reported here is not focused on fabricating transparent coatings, but on increasing the absolute EMI SE values of these coatings. Our method allows us, thus, to fabricate bulk (*i.e.* non-transparent) structures but still thin (up to 30 μm thick) and highly ordered graphene based coatings to maximize the electrical conductivity and EMI SE with the minimum possible thickness and amount of graphene employed. This underlines the importance of designing highly-ordered structures with optimized electrical properties in order to maximize the EMI shielding, minimizing both thickness and graphene content.

CONCLUSIONS

We report here a scalable spray coating layer-by-layer approach to fabricate thin, highly-ordered and electrically-conductive multilayer coatings for EMI shielding, through alternating self-assembly of negatively-charged reduced graphene oxide (RGO) and a positively-charged polyelectrolyte *n* times on a substrate to fabricate (PEI/RGO)$_n$ coatings. SEM micrographs of the transverse sections of the LbL coatings revealed highly-ordered and well-defined multi-layered microstructures, further confirmed by polarized Raman spectroscopy, with thicknesses depending strongly on the loading of graphene per bi-layer and *n*. Percolation theory was applied to these structures in order to understand the conduction mechanism. Outstanding EMI shielding performances were found for these coating structures through a mechanism of absorption. Maximum values of EMI shielding effectiveness ~29 dB for 6 μm thick coatings with 19 vol.% loading of reduced electrochemically-produced graphene oxide were found, which are considerably higher than those values previously reported for thicker graphene papers and graphene/polymer composite films with higher graphene contents, making our coatings potentially promising for EMI shielding applications in sectors where high EMI SE is important (especially if EMI absorbance is required) whilst transparency is not needed. In addition, we probed here that the EMI shielding performance of graphene based structures can be tuned by varying the nature and electrical properties of the graphene employed.

EXPERIMENTAL SECTION

**Materials**

Polyethyleneimine (PEI) solution ~50% in $H_2O$ was purchased from Sigma-Aldrich. The graphene materials used in this work have been prepared as described in the next section.

**Preparation and characterization of TRGO and R-EGO**

In order to remove oxygen-containing functionalities from the surface of the flakes, GO powder (prepared using Hummers' method) was thermally reduced in a nitrogen atmosphere by a rapid heating up to 750 °C using a metallic reactor heated with a gas burner[53]. TRGO was obtained as a black powder of very low bulk density. The super heating is the prime requirement to achieve exfoliation of graphene sheets simultaneously to the reduction. In order to control the diameter of the TRGO, two types of starting graphite were employed. A micro-graphite with a particle size distribution $d_{50}$ = 17-23 μm (SGA 20 M 99.5; AMG graphite, Hauzenberg, Germany) to prepare

TRGO with a particle size distribution between 5-200 µm and an average diameter of ~20 µm (named 20µm-TRGO). Graphite with larger particles (Rfl 99.5; AMG graphite, Hauzenberg, Germany; min. 90% > 160 µm) was employed to produce TRGO with a particle size between 5-900 µm and $d_{50}$ ~100 µm (named 100µm-TRGO).

The combination of different average diameters (~20 µm vs. ~100 µm characteristic of 20µm-TRGO and 100µm-TRGO flakes, respectively, as revealed by TEM, SI) and similar thicknesses (between 3.5 nm and 10 nm as revealed by AFM, SI), make these materials ideal to evaluate the role of lateral flake dimensions on microstructure and properties of the LbL coatings. (Further characterization of the TRGO filler materials, including XPS showing an effective reduction, is shown in SI).

EGO was prepared following the experimental procedure reported elsewhere.[52] For the EGO based LbL coatings, the reduction was performed after deposition (as described below).

**Preparation of the LbL structures**

Multi-layered graphene coatings are prepared using a LbL approach by alternating self-assembly of negatively charged TRGO and positively-charged PEI, *n* times on a PET substrate, as represented schematically in Figure 1. Zeta potential revealed negative charges of about -50 mV and -40 mV for dispersions of 100µm-TRGO and 20µm-TRGO, respectively, in ethanol (0.5 mg/mL), whereas a positive charge of about 12 mV was found for the PEI/$H_2O$ solution (10 mg/mL). A commercial aerosol (Linden$^{TM}$ H-BADG-AIRBRUSH-KIT) was used to deposit graphene and PEI by spraying the dispersions of TRGO in ethanol and the aqueous solutions of PEI, alternatively, on a PET substrate. The coatings were dried using an air flow at room temperature after each deposition. The effect of rinsing after each deposition was evaluated and is discussed in the SI.

In order to assess the formation of the LbL system, the amount of graphene deposited per cycle (*i.e.* mg of graphene per deposition cycle and unit surface area, defined here as 'areal mass of graphene') is varied whilst the areal mass of PEI is kept constant (= 0.18 mg/cycle/cm$^2$). The areal masses of TRGO and PEI are used to calculate the loading of graphene in the LbL structures as a weight per unit volume fraction. The loading depends on the areal masses of graphene and PEI (consequently, 'graphene loading per cycle' is equivalent to just 'graphene loading'), but is independent of $n$. In order to assess the interaction between layers, $n$ is varied between 1 and 10. The graphene loading, *i.e.* the amount of graphene deposited per cycle can be controlled accurately by modifying the deposited volume of a dispersion of graphene with a known concentration. In a similar way, the final thickness of the fabricated coatings can be also controlled accurately through controlling the graphene deposited per cycle and $n$.

For the preparation of the R-EGO based LbL coatings, the EGO (prepared following the method reported elsewhere)[52] was first deposited from a 1 mg/mL aqueous dispersion (which shows a negative charge of around -45 mV, as measured by Zeta potential). After deposition, the LbL EGO coatings were placed in a sealed bottle (100 mL in volume), containing 0.5 ml of HI (55%) and 1.25 mL of acetic acid, in a water bath at 95 °C for 1 h for reduction. The reduced EGO (R-EGO) coatings were rinsed with ethanol four times to remove any residual acid and iodine and dried with N$_2$ gas. An effective reduction of the EGO materials prepared here is shown by Raman analysis (Figure S7, SI), along with further details on the reduction procedure.

**Characterization**

A Zeiss Ultra 55 FEG-SEM (scanning electronic microscope) is employed to analyse the surface morphology and the transversal section of the LbL coatings (using a EHT of 2 kV). The orientation of the graphene flakes is evaluated by polarized Raman spectroscopy using a backscattering geometry and a VV (vertical/vertical) combination of incident and scattered polarization, in which the directions of the incident and scattered polarization were the same. A rotational stage is employed to rotate the specimen with respect to its axes. The change in $I_D$ is recorded as a function of rotation angle, following the method by Li *et al.*[38].

The sample thicknesses are measured with a digital micrometer IP65 QUANTUMIKE. The impedance is tested on 2 mm x 3 mm specimens using a PSM 1735 Frequency Response Analyzer from Newtons4th Ltd connected with Impedance Analysis Interface (IAI) at the range of frequencies from 1 to $10^6$ Hz. The conductivities ($\sigma$) of the coatings are calculated from the measured impedances as:[40]

$$\sigma(\omega) = |Y^*(\omega)|\frac{t}{A} = \frac{1}{Z^*} \times \frac{t}{A} \qquad \text{(Equation (3))}$$

where $Y^*(\omega)$ is the complex admittance, $Z^*$ is the complex impedance, $t$ and $A$ are the thickness and cross section area of the sample.

An Evolution 201 Thermo Scientific UV-Vis spectrophotometer is employed to analyse the absorption and transmittance in the range 300-800 nm. EMI shielding tests are performed using a PNA (vector network analyser 10MHz- 50 GHz), 2 RF cables and one pair of waveguides in the X-band frequency range (8.4 - 12.4 GHz). The samples (2.5 x 1.5 cm coatings) are compressed in the middle of two waveguides while VNA measured the $S_{21}$ (transmittance), from which SE is calculated.

## ASSOCIATED CONTENT

**Supporting Information**.

The following files are available free of charge.

Details of fabrication, characterization and calculations (PDF).

## AUTHOR INFORMATION

**Corresponding Authors**

* E-mail: cristina.valles@manchester.ac.uk

* E-mail: ian.kinloch@manchester.ac.uk

**Author Contributions**

The manuscript was written through contributions of all authors. All authors have given approval to the final version of the manuscript.

**Notes**

The authors declare no competing financial interest.

## ACKNOWLEDGMENT

This project has received funding from the European Union's Horizon 2020 research and innovation programme under grant agreement No 785219.

## REFERENCES

(1)    Shahzad, F.; Alhabeb, M.; Hatter, C. B.; Anasori, B.; Hong, S. M.; Koo, C. M.; Gogotsi, Y., Electromagnetic interference shielding with 2D transition metal carbides (MXenes). *Science* 2016, 353, 1137-1140.

(2)     Chen, Z. P.; Xu, C.; Ma, C. Q.; Ren, W. C.; Cheng, H. M., Lightweight and Flexible Graphene Foam Composites for High-Performance Electromagnetic Interference Shielding. *Adv Mater* 2013, 25, 1296-1300.

(3)     Yan, D. X.; Pang, H.; Li, B.; Vajtai, R.; Xu, L.; Ren, P. G.; Wang, J. H.; Li, Z. M., Structured Reduced Graphene Oxide/Polymer Composites for Ultra-Efficient Electromagnetic Interference Shielding. *Adv Funct Mater* 2015, 25, 559-566.

(4)     Yousefi, N.; Gudarzi, M. M.; Zheng, Q.; Lin, X.; Shen, X.; Jia, J.; Sharif, F.; Kim, J. K., Highly aligned, ultralarge-size reduced graphene oxide/polyurethane nanocomposites: Mechanical properties and moisture permeability. *Composites Part A: Applied Science and Manufacturing* 2013, 49, 42-50.

(5)     Zhang, Y.; Huang, Y.; Zhang, T. F.; Chang, H. C.; Xiao, P. S.; Chen, H. H.; Huang, Z. Y.; Chen, Y. S., Broadband and Tunable High-Performance Microwave Absorption of an Ultralight and Highly Compressible Graphene Foam. *Adv Mater* 2015, 27, 2049-2053.

(6)     Yang, Y. L.; Gupta, M. C., Novel carbon nanotube-polystyrene foam composites for electromagnetic interference shielding. *Nano Lett* 2005, 5, 2131-2134.

(7)     Chung, D. D. L., Electromagnetic interference shielding effectiveness of carbon materials. *Carbon* 2001, 39, 279-285.

(8)     Geetha, S.; Kumar, K. K. S.; Rao, C. R. K.; Vijayan, M.; Trivedi, D. C., EMI Shielding: Methods and Materials-A Review. *J Appl Polym Sci* 2009, 112, 2073-2086.

(9)     Zhang, H. B.; Yan, Q.; Zheng, W. G.; He, Z. X.; Yu, Z. Z., Tough Graphene-Polymer Microcellular Foams for Electromagnetic Interference Shielding. *Acs Appl Mater Inter* 2011, 3, 918-924.

(10)     Eswaraiah, V.; Sankaranarayanan, V.; Ramaprabhu, S., Functionalized Graphene-PVDF Foam Composites for EMI Shielding. *Macromol Mater Eng* 2011, 296, 894-898.

(11)     Fletcher, A.; Gupta, M. C.; Dudley, K. L.; Vedeler, E., Elastomer foam nanocomposites for electromagnetic dissipation and shielding applications. *Compos Sci Technol* 2010, 70, 953-958.

(12)     Thomassin, J. M.; Pagnoulle, C.; Bednarz, L.; Huynen, I.; Jerome, R.; Detrembleur, C., Foams of polycaprolactone/MWNT nanocomposites for efficient EMI reduction. *J Mater Chem* 2008, 18, 792-796.

(13)     Li, N.; Huang, Y.; Du, F.; He, X. B.; Lin, X.; Gao, H. J.; Ma, Y. F.; Li, F. F.; Chen, Y. S.; Eklund, P. C., Electromagnetic interference (EMI) shielding of single-walled carbon nanotube epoxy composites. *Nano Lett* 2006, 6, 1141-1145.

(14)     Liang, J. J.; Wang, Y.; Huang, Y.; Ma, Y. F.; Liu, Z. F.; Cai, J. M.; Zhang, C. D.; Gao, H. J.; Chen, Y. S., Electromagnetic interference shielding of graphene/epoxy composites. *Carbon* 2009, 47, 922-925.

(15)     Liu, Z. F.; Bai, G.; Huang, Y.; Ma, Y. F.; Du, F.; Li, F. F.; Guo, T. Y.; Chen, Y. S., Reflection and absorption contributions to the electromagnetic interference shielding of single-walled carbon nanotube/polyurethane composites. *Carbon* 2007, 45, 821-827.

(16)     Wang, L. L.; Tay, B. K.; See, K. Y.; Sun, Z.; Tan, L. K.; Lua, D., Electromagnetic interference shielding effectiveness of carbon-based materials prepared by screen printing. *Carbon* 2009, 47, 1905-1910.

(17)     Yang, Y. L.; Gupta, M. C.; Dudley, K. L.; Lawrence, R. W., Conductive carbon nanoriber-polymer foam structures. *Adv Mater* 2005, 17, 1999-+.

(18) Park, S.; Ruoff, R. S., Chemical methods for the production of graphenes. *Nat Nanotechnol* 2009, 4, 217-224.

(19) Rao, C. N. R.; Sood, A. K.; Subrahmanyam, K. S.; Govindaraj, A., Graphene: The New Two-Dimensional Nanomaterial. *Angew Chem Int Edit* 2009, 48, 7752-7777.

(20) Ferrari, A. C.; Bonaccorso, F.; Fal'ko, V.; Novoselov, K. S.; Roche, S.; Boggild, P.; Borini, S.; Koppens, F. H. L.; Palermo, V.; Pugno, N.; Garrido, J. A.; Sordan, R.; Bianco, A.; Ballerini, L.; Prato, M.; Lidorikis, E.; Kivioja, J.; Marinelli, C.; Ryhanen, T.; Morpurgo, A. Coleman J.N., Nicolosi V., Colombo L., Fert A., Garcia-Hernandez M., Bachtold A., Schneider G.F., Guinea F., Dekker C., Barbone M., Sun Z., Galiotis C., Grigorenko A.N., Konstantatos G., Kis A., Katsnelson M., Vandersypen L., Loiseau A., Morandi V., Neumaier D., Treossi E., Pellegrini V., Polini M., Tredicucci A., Williams G.M, Hong B.H., Ahn J.H., Kim J.M., Zirath H., van Wees B.J., van der Zant H., Occhipinti L., Di Matteo A., Kinloch I.A., Seyller T., Quesnel E., Feng X., Teo K., Rupesinghe N., Hakonen P., Neil S.R.T., Tannock Q., Löfwanderaq T. and Kinaretba J. Science and technology roadmap for graphene, related two-dimensional crystals, and hybrid systems. *Nanoscale* 2015, 7, 4598-4810.

(21) Vallés, C.; Young, R. J.; Lomax, D. J.; Kinloch, I. A., The rheological behaviour of concentrated dispersions of graphene oxide. *J Mater Sci* 2014, 49, 6311-6320.

(22) Wan, Y. J.; Zhu, P. L.; Yu, S. H.; Sun, R.; Wong, C. P.; Liao, W. H., Ultralight, super-elastic and volume-preserving cellulose fiber/graphene aerogel for high-performance electromagnetic interference shielding. *Carbon* 2017, 115, 629-639.

(23) Wan, Y. J.; Zhu, P. L.; Yu, S. H.; Sun, R.; Wong, C. P.; Liao, W. H., Graphene paper for exceptional EMI shielding performance using large-sized graphene oxide sheets and doping strategy. *Carbon* 2017, 122, 74-81.

(24) Wang, Z. C.; Wei, R. B.; Liu, X. B., Fluffy and Ordered Graphene Multilayer Films with Improved Electromagnetic Interference Shielding over X-Band. *Acs Appl Mater Inter* 2017, 9, 22408-22419.

(25) Jan, R.; Habib, A.; Akram, M. A.; Ahmad, I.; Shah, A.; Sadiq, M.; Hussain, A., Flexible, thin films of graphene-polymer composites for EMI shielding. *Mater Res Express* 2017, 4.

(26) Song, W. L.; Cao, M. S.; Lu, M. M.; Bi, S.; Wang, C. Y.; Liu, J.; Yuan, J.; Fan, L. Z., Flexible graphene/polymer composite films in sandwich structures for effective electromagnetic interference shielding. *Carbon* 2014, 66, 67-76.

(27) Jiang, Q.; Liao, X.; Li, J.; Chen, J.; Wang, G.; Yi, J.; Yang, Q.; Li, G., Flexible thermoplastic polyurethane/reduced graphene oxide composite foams for electromagnetic interference shielding with high absorption characteristic. *Composites Part A: Applied Science and Manufacturing* 2019, 123, 310-319.

(28) Gupta, S.; Sharma, S. K.; Pradhan, D.; Tai, N. H., Ultra-light 3D reduced graphene oxide aerogels decorated with cobalt ferrite and zinc oxide perform excellent electromagnetic interference shielding effectiveness. *Composites Part A: Applied Science and Manufacturing* 2019, 123, 232-241.

(29) Zhang, N.; Wang, Z.; Song, R.; Wang, Q.; Chen, H.; Zhang, B.; Lv, H.; Wu, Z.; He, D., Flexible and transparent graphene/silver-nanowires composite film for high electromagnetic interference shielding effectiveness. *Science Bulletin* 2019, 64, 540-546.

(30) Lin, S. F.; Ju, S.; Shi, G.; Zhang, J. W.; He, Y. L.; Jiang, D. Z., Ultrathin nitrogen-doping graphene films for flexible and stretchable EMI shielding materials. *J Mater Sci* 2019, 54, 7165-7179.

(31)     Decher, G.; Hong, J. D.; Schmitt, J., Buildup of Ultrathin Multilayer Films by a Self-Assembly Process .3. Consecutively Alternating Adsorption of Anionic and Cationic Polyelectrolytes on Charged Surfaces. *Thin Solid Films* 1992, 210, 831-835.

(32)     Yang, Y. H.; Bolling, L.; Priolo, M. A.; Grunlan, J. C., Super Gas Barrier and Selectivity of Graphene Oxide-Polymer Multilayer Thin Films. *Adv Mater* 2013, 25, 503-508.

(33)     Stevens, B.; Dessiatova, E.; Hagen, D. A.; Todd, A. D.; Bielawski, C. W.; Grunlan, J. C., Low-Temperature Thermal Reduction of Graphene Oxide Nanobrick Walls: Unique Combination of High Gas Barrier and Low Resistivity in Fully Organic Polyelectrolyte Multilayer Thin Films. *Acs Appl Mater Inter* 2014, 6, 9942-9945.

(34)     Pierleoni, D.; Minelli, M.; Ligi, S.; Christian, M.; Funke, S.; Reineking, N.; Morandi, V.; Doghieri, F.; Palermo, V., Selective Gas Permeation in Graphene Oxide-Polymer Self-Assembled Multilayers. *Acs Appl Mater Inter* 2018, 10, 11242-11250.

(35)     Kim, S.; Oh, J. S.; Kim, M. G.; Jang, W.; Wang, M.; Kim, Y.; Seo, H. W.; Kim, Y. C.; Lee, J. H.; Lee, Y.; Nam, J. D., Electromagnetic Interference (EMI) Transparent Shielding of Reduced Graphene Oxide (RGO) Interleaved Structure Fabricated by Electrophoretic Deposition. *Acs Appl Mater Inter* 2014, 6, 17647-17653.

(36)     Lu, Z. G.; Ma, L. M.; Tan, J. B.; Wang, H. Y.; Ding, X. M., Transparent multi-layer graphene/polyethylene terephthalate structures with excellent microwave absorption and electromagnetic interference shielding performance. *Nanoscale* 2016, 8, 16684-16693.

(37)     Kang, J.; Kim, D.; Kim, Y.; Choi, J. B.; Hong, B. H.; Kim, S. W., High-performance near-field electromagnetic wave attenuation in ultra-thin and transparent graphene films. *2D Mater* 2017, 4.

(38) Li, Z.; Young, R. J.; Kinloch, I. A., Interfacial stress transfer in graphene oxide nanocomposites. *ACS Applied Materials and Interfaces* 2013, 5, 456-463.

(39) Casiraghi, C.; Hartschuh, A.; Qian, H.; Piscanec, S.; Georgi, C.; Fasoli, A.; Novoselov, K. S.; Basko, D. M.; Ferrari, A. C., Raman Spectroscopy of Graphene Edges. *Nano Lett* 2009, 9, 1433-1441.

(40) Stauffer D, A. A., Introduction to percolation theory. *London: Taylor and Francis* 1994, 175.

(41) Chatterjee, S.; Nafezarefi, F.; Tai, N. H.; Schlagenhauf, L.; Nüesch, F. A.; Chu, B. T. T., Size and synergy effects of nanofiller hybrids including graphene nanoplatelets and carbon nanotubes in mechanical properties of epoxy composites. *Carbon* 2012, 50, 5380-5386.

(42) Pike GE, S. C., Percolation and conductivity: A computer study. I*. *Physical Review B* 1974, 10, 14.

(43) Gao, J. F.; Li, Z. M.; Meng, Q. J.; Yang, Q., CNTs/UHMWPE composites with a two-dimensional conductive network. *Mater Lett* 2008, 62, 3530-3532.

(44) Zhao, S. A.; Chang, H. Y.; Chen, S. J.; Cui, J.; Yan, Y. H., High-performance and multifunctional epoxy composites filled with epoxide-functionalized graphene. *Eur Polym J* 2016, 84, 300-312.

(45) Du, J. H.; Zhao, L.; Zeng, Y.; Zhang, L. L.; Li, F.; Liu, P. F.; Liu, C., Comparison of electrical properties between multi-walled carbon nanotube and graphene nanosheet/high density polyethylene composites with a segregated network structure. *Carbon* 2011, 49, 1094-1100.

(46) Wang, D. G.; Wang, X. G., Self-Assembled Graphene/Azo Polyelectrolyte Multilayer Film and Its Application in Electrochemical Energy Storage Device. *Langmuir* 2011, 27, 2007-2013.

(47) Wang, L.; Wang, W. C.; Fu, Y.; Wang, J. J.; Lvov, Y.; Liu, J.; Lu, Y. L.; Zhang, L. Q., Enhanced electrical and mechanical properties of rubber/graphene film through layer-by-layer electrostatic assembly. *Compos Part B-Eng* 2016, 90, 457-464.

(48) Ambrosetti, G.; Johner, N.; Grimaldi, C.; Danani, A.; Ryser, P., Percolative properties of hard oblate ellipsoids of revolution with a soft shell. *Phys Rev E* 2008, 78.

(49) Li, J.; Kim, J. K., Percolation threshold of conducting polymer composites containing 3D randomly distributed graphite nanoplatelets. *Compos Sci Technol* 2007, 67, 2114-2120.

(50) Marsden A.J, P. D. G., Valles C, Liscio A, Palermo V, Bissett M.A, Young R.J, Kinloch I.A., Electrical percolation in graphene-polymer composites. *2d Mater* 2018, 5.

(51) Huang, H. D.; Liu, C. Y.; Zhou, D.; Jiang, X.; Zhong, G. J.; Yan, D. X.; Li, Z. M., Cellulose composite aerogel for highly efficient electromagnetic interference shielding. *J Mater Chem A* 2015, 3, 4983-4991.

(52) Cao, J. Y.; He, P.; Mohammed, M. A.; Zhao, X.; Young, R. J.; Derby, B.; Kinloch, I. A.; Dryfe, R. A. W., Two-Step Electrochemical Intercalation and Oxidation of Graphite for the Mass Production of Graphene Oxide. *J Am Chem Soc* 2017, 139, 17446-17456.

(53) Steurer, P.; Wissert, R.; Thomann, R.; Mülhaupt, R., Functionalized graphenes and thermoplastic nanocomposites based upon expanded graphite oxide. *Macromol Rapid Comm* 2009, 30, 316-327.

For Table of Contents Only

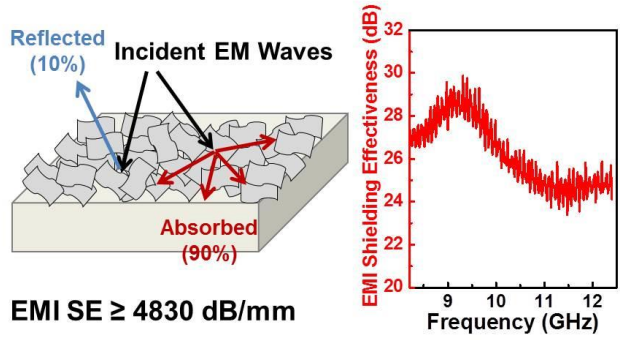